\documentstyle[epsfig,preprint,aps]{revtex}
\def\bge{\begin{equation}}
\def\ene{\end{equation}}
\def\bg{\begin{eqnarray}}
\def\en{\end{eqnarray}}

\def\S0{{\Sigma^0}}

\def\k0bar{\bar{K}^0}

\def\vr{\vec{r}}
\def\vx{\vec{x}}

\begin{document}
\tighten
\draft
\begin{titlepage}
\title{Charge symmetry breaking in mirror nuclei from quarks}
\author{
K. Tsushima$^1$~\thanks{ktsushim@physics.adelaide.edu.au}~,
K. Saito$^2$~\thanks{ksaito@nucl.phys.tohoku.ac.jp}~, 
A. W. Thomas$^1$~\thanks{athomas@physics.adelaide.edu.au}}  
\address{$^1$Special Research Center for 
the Subatomic Structure of Matter\break
and Department of Physics and Mathematical Physics\break 
University of Adelaide, SA 5005, Australia} 
\address{$^2$Tohoku College of Pharmacy, Sendai 981-8558, Japan} 
\maketitle
\vspace{-5.5cm}
\hfill ADP-99-28/T367
\vspace{5.5cm}
\begin{abstract}
The binding energy differences of the valence proton and 
neutron of the mirror nuclei, $^{15}$O -- $^{15}$N, $^{17}$F -- $^{17}$O,
$^{39}$Ca -- $^{39}$K and $^{41}$Sc -- $^{41}$Ca, 
are calculated using the quark-meson coupling 
(QMC) model. The calculation involves nuclear structure 
and shell effects explicitly. It is shown that 
binding energy differences of a few hundred keV arise from the strong 
interaction, even after subtracting all electromagnetic corrections. 
The origin of these differences may be ascribed to the charge symmetry 
breaking effects set in the strong interaction through 
the u and d current quark mass difference.
\end{abstract}
\end{titlepage}
%

The discrepancy between the calculated binding energy differences 
of mirror nuclei and those measured is a long-standing problem 
in nuclear physics. It is known as the   
Okamoto-Nolen-Schiffer (ONS) anomaly~\cite{oka,nor,shl}.
Although it was first thought that electromagnetic effects 
could almost account for the observed binding energy differences, 
it is now believed that the ONS anomaly has its origin in 
charge symmetry breaking (CSB) in the strong interaction~\cite{miller}.
In addition to calculations based on charge symmetry violating 
meson exchange potentials \cite{miller,blu,suz,krein,shah},
a number of quark-based calculations  
have been performed~\cite{wil,hen,hatsuda,adami,fio,saito,sch} 
in an attempt to resolve this anomaly.   
To some extent, these have been stimulated by the discovery 
of the nuclear dependence of the nucleon 
structure function measured in deep inelastic lepton-nucleus 
scattering (the nuclear EMC effect~\cite{emc}). 
In such calculations, CSB enters 
through the up (u) and down (d) current quark mass difference in QCD.
Despite these efforts, the difficulty of producing a realistic description
of nuclear 
structure on the basis of explicit quark degrees of freedom has  
hindered the direct calculation of the binding energy differences.
(One such quark-based, nuclear calculation exists~\cite{nak}, but it  
involved a shell model calculation for the iso-vector mass shifts 
of iso-spin multiplets in $1s0d$-shell nuclei, and the role of quarks  
entered through a model for the short-range CSB force.)

In this study we report the results for the binding energy differences 
of the valence (excess) proton and neutron of the mirror nuclei, 
$^{15}$O -- $^{15}$N, $^{17}$F -- $^{17}$O, $^{39}$Ca -- $^{39}$K and
$^{41}$Sc -- $^{41}$Ca, calculated using a quark-based model    
involving explicit nuclear structure and shell   
effects -- the quark-meson coupling (QMC) model~\cite{finite0,finite1}.
This model has been successfully applied not only to 
traditional nuclear problems~\cite{finite0,finite1,hyper}  
but also to other new areas as well~\cite{tony}.
Although some exploratory QMC results on the ONS anomaly   
have already been reported~\cite{saito}, an early version of the model was 
used there, and it was applied to finite nuclei only through local 
density approximation, rather than a consistent shell model calculation.

A detailed description of the Lagrangian density and the
mean-field equations of motion needed to describe a finite nucleus
is given in Refs.~\cite{finite0,finite1}.
A major difference in the present work compared with 
Refs.~\cite{finite0,finite1} 
is that here charge symmetry is explicitly broken at the 
quark level through their masses. 
We use different values for the u and d current quark masses, 
and the proton and neutron (effective) masses.
Thus, the saturation properties of symmetric nuclear matter needed to be 
recalculated to fix the relevant quark-meson coupling constants. 
At position $\vr$ in a
nucleus (the coordinate origin is taken at the center of the nucleus),
the Dirac equations for the quarks in the proton or 
neutron bag are given by:
$$
\left[ i \gamma \cdot \partial_x - \left(
\left(\begin{array}{c} m_u\\ m_d\\ \end{array}\right)
 - V^q_\sigma(\vr) \right)
- \gamma^0
\left( V^q_\omega(\vr) \pm \frac{1}{2} V^q_\rho(\vr) \right) \right]
\left(\begin{array}{c} \psi_u(x)\\ \psi_d(x)\\ \end{array}\right) = 0, 
\nonumber
$$
\bge 
\hspace*{20em}(|\vx - \vr| \le {\rm bag\,\, radius}). 
\label{diraceq}
\ene
The mean-field potentials for a bag centered at position $\vr$ in
the nucleus are defined by $V^q_\sigma(\vr) = g^q_\sigma
\sigma(\vr), V^q_\omega(\vr) = g^q_\omega \omega(\vr)$ and
$V^q_\rho(\vr) = g^q_\rho b(\vr)$, with $g^q_\sigma, g^q_\omega$ and
$g^q_\rho$ the corresponding quark and meson-field coupling
constants. (Note that we have neglected a possible, very slight
variation of the scalar and vector mean-fields inside the nucleon bag
due to its finite size~\cite{finite0}.)  
The mean meson fields are  
calculated self-consistently by solving Eqs.~(23) -- (30) of
Ref.~\cite{finite1} with the proper modifications caused by 
the different proton  
and neutron (u and d quark) masses, namely, by solving a set of coupled, 
non-linear, differential equations for static, spherically symmetric nuclei,
resulting from the variation of the effective Lagrangian density
involving the quark degrees of freedom and the scalar, vector and Coulomb
fields in mean field approximation.
Thus, the present calculation is free from the sort of double counting 
questioned by Auerbach~\cite{aue}, namely 
that one should not add the effective mass 
difference reduction between the proton and neutron in medium on the 
top of the Coulomb displacement energies. 
Furthermore, the calculation also includes the shell effects which were 
discussed by Cohen et al.~\cite{coh}. 

Before discussing the results obtained, we need to specify the  
parameters and inputs used in the calculation. 
They are summarized in TABLE~\ref{parameters}.
The bag constant, $B$, is determined by the bare proton mass in free space 
after allowing for the electromagnetic self-energy correction, $+$0.63 MeV. 
The parameter $z$ represents the sum of the 
center-of-mass and gluon fluctuation
corrections, included in the standard MIT bag mass formula as  
$-z/R$ and assumed independent of the density~\cite{finite0}. 
$B$ and $z$ are determined by setting the bag radius in free space
to be $R = 0.8$ fm,
and imposing the mass stability condition, 
$\frac{\partial m_p}{\partial R} = 0$~\cite{finite0}.
(See Ref.~\cite{finite0} for details.)
For the neutron, the procedure is the same as that for the proton, 
allowing for the electromagnetic self-energy 
correction, $-$0.13 MeV, but using the values of $B$ and $z$ 
determined above and 
calculating the d current quark mass and the bag radius for the   
neutron by the mass stability condition.
Thus, the u current quark mass, $m_u$, is the basic input parameter 
used to fix the model parameters so as to reproduce the proton and 
neutron masses in free space after allowing for the electromagnetic 
self-energy corrections.
The coupling constants, $g^q_\sigma$ and $g^q_\omega$ are 
determined so as to fit the saturation properties of 
symmetric nuclear matter -- i.e., 
a binding energy of 15.7 MeV at the saturation 
density, $\rho_0 = 0.15$ fm$^{-3}$. The binding energy is calculated 
by subtracting the average nucleon mass, $(m_p + m_n)/2$, 
and using the different scalar densities for protons and neutrons. 
In TABLE~\ref{parameters}, SU(2) stands for the parameters and inputs 
obtained and used for the calculation when SU(2) symmetry for the quarks  
and nucleons is assumed -- i.e., using the same values for the u and d 
quark masses, and also for the proton and neutron masses.
We should notice that the coupling constant, $g^q_\sigma$, 
is also scaled for the present calculation of finite nuclei, by the same  
amount as found necessary in Ref.~\cite{finite1} to fit the r.m.s. 
charge radius of $^{40}$Ca (for the bag radius 0.8 fm) -- 
keeping the ratio $(g_\sigma/m_\sigma)$ fixed,  
because the fixed ratio of $(g_\sigma/m_\sigma)$ has no effect 
on the properties of infinite nuclear matter. 

The quark-$\rho$ meson coupling constant, $g^q_\rho$, needs 
some explanation. Within QMC, $g^q_\rho$ is determined so as to 
reproduce the symmetry energy of 35 MeV. However, because the present 
model does not contain the $\rho$-nucleon tensor 
coupling~\cite{finite0,finite1} and we work only in Hartree 
approximation~\cite{gaston}, this  
gives an unrealistically large value for the coupling constant, 
$g^q_\rho = 9.321$. 
To make a realistic estimate, taking into account the $\rho$-meson 
central and spin-orbit potentials for the valence proton and  
neutron binding energies, we use the phenomenological
value, $g^q_\rho = 4.595$ ($g^2_{\rho NN}/4\pi = 4 \times 0.42$), the value 
at zero three-momentum transfer corresponding to Hartree approximation, 
from TABLE 4.1 of Ref.~\cite{mach}.
We also estimate the contributions of the $\rho$-potentials 
using the naive QMC value, $g^q_\rho = 9.321$, 
in order to test the sensitivity.

\begin{figure}[hbt]
\begin{center}
\epsfig{file=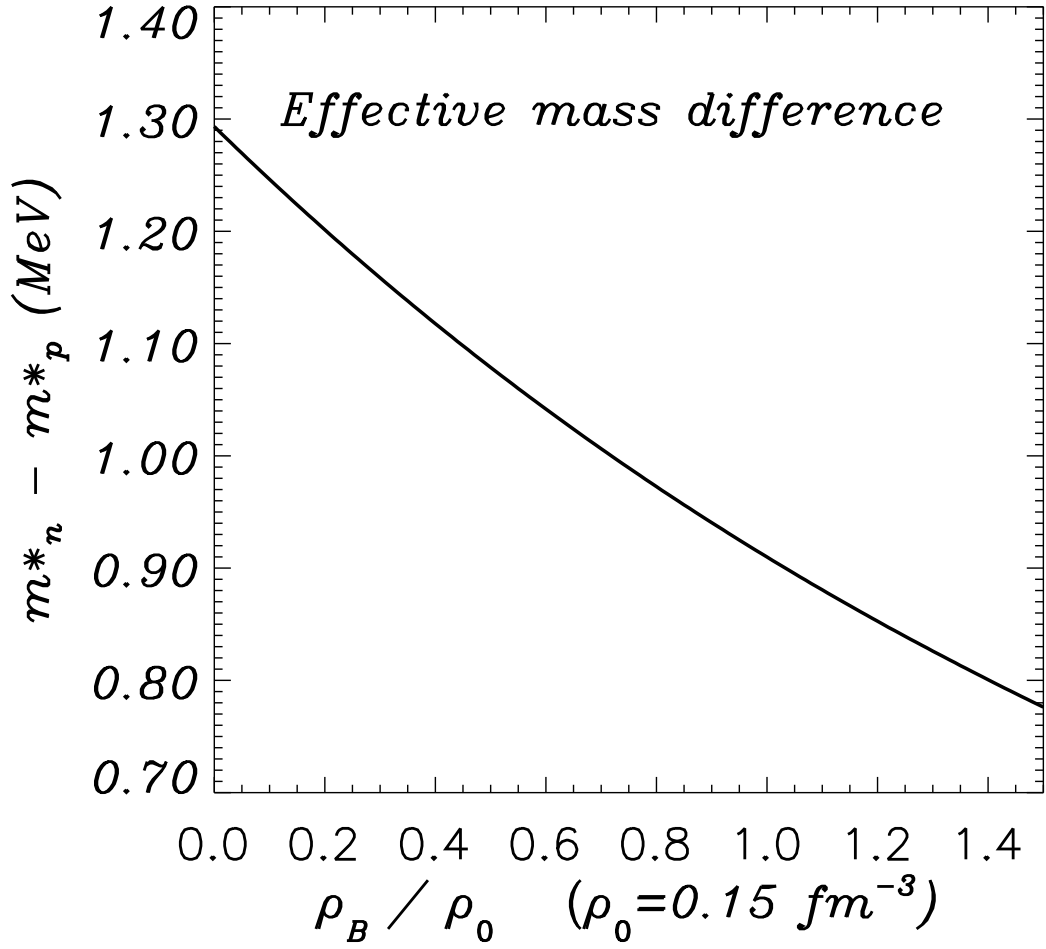,height=14cm}
\caption{Proton-neutron effective mass difference in symmetric nuclear matter 
with the electromagnetic self-energy corrections.}
\label{onsmass}
\end{center}
\end{figure}

In FIG.~\ref{onsmass} we show the proton and neutron effective mass
difference calculated in symmetric nuclear matter, including 
the electromagnetic self-energy corrections for the proton ($+$0.63 MeV) and
neutron ($-$0.13 MeV). 
One notices that the proton and neutron mass difference
becomes smaller as the baryon density increases -- a result which 
was also found in Refs.~\cite{hen,hatsuda,adami,saito}. 
This seems to work in the right direction to resolve the ONS anomaly, 
but it is still not quantitative.
(Recall the discussion of Auerbach~\cite{aue}.)

Next, we show the calculated single-particle energies 
for $^{17}$F and $^{17}$O in TABLE~\ref{energylevel}, as an example.  
These mirror nuclei 
have a common core nucleus, $^{16}$O, and have an extra valence 
proton for $^{17}$F and neutron for $^{17}$O.
In order to focus on the strong interaction effect  
for the valence proton and neutron,  
the Dirac equations are solved without the 
Coulomb and $\rho$-meson potentials, or 
the electromagnetic self-energy corrections, and keeping 
only the charge symmetric $\sigma$ and $\omega$ 
mean field potentials. 
Consistently, the valence nucleon contributions are not included in the 
Coulomb (proton) and $\rho$-mean field (iso-vector) source densities 
in the core nucleus.
However, for the nucleons in the core nucleus, electromagnetic self-energy 
corrections and the Coulomb potential as well as the $\rho$ mean field 
potential are included in addition
to the $\sigma$ and $\omega$ mean field potentials in solving 
the Dirac equations.
Results will be shown for three cases for $^{17}$F and $^{17}$O: 
\begin{enumerate}
\item Calculation performed imposing 
charge symmetry breaking through the u and d quark masses 
and the proton and neutron masses  
using the phenomenological $\rho$-quark coupling constant, 
$g^q_\rho = 4.595$ (at zero three-momentum transfer, $\vec{q}=0$, 
corresponding to Hartree approximation~\cite{mach}) (denoted CSB).
\item Calculation performed assuming 
SU(2) symmetry for the u and d quark masses and the proton and neutron   
masses using the phenomenological $\rho$-quark coupling constant, 
$g^q_\rho = 4.595$ (denoted SU(2)). (See also the explanation of CSB.)
\item Calculation performed 
imposing charge symmetry breaking through the different 
u and d current quark masses 
and the proton and neutron masses  
using the $\rho$-quark coupling constant, 
$g^q_\rho = 9.321$ (denoted Case 3). 
\end{enumerate}

The SU(2) results for $^{17}$F and $^{17}$O agree perfectly  
with each other as they should.   
Single-particle energies in the cores of $^{17}$F and 
$^{17}$O are slightly different for both CSB and Case 3. 
This difference is induced by the different (effective) masses for 
the valence proton and neutron, arising from the charge and density
dependence of their coupling to the self-consistent scalar mean field.
This also causes a second order effect on the Coulomb and $\rho$-meson 
potentials through the self-consistency procedure.
The single-particle energies of the valence proton and neutron 
are practically equal for both CSB and Case 3.

It is interesting to compare the binding energy differences  
between the valence proton in $^{17}$F and neutron in $^{17}$O. 
Both CSB and Case 3 results give,  
$E(p)(1d_{5/2}) - E(n)(1d_{5/2}) \simeq 0.18$ MeV, while the SU(2) case 
is zero as it should be. 
This amount already shows a magnitude similar to that of the 
observed binding energy differences, where the origin may be 
ascribed to the effect of the proton-neutron effective mass difference 
reduction and simultaneous effect of the core nucleus potentials. 

Note that in QMC the quark scalar charge for the d quark,
which is defined by the integral of the quark scalar density over the
nucleon volume, is slightly greater than that for the u quark, because
the u quark mass is smaller than the d quark mass. 
The lower component of the u quark wave function  
is enhanced more than that of the d quark.
This is a simple consequence of relativistic quantum mechanics. 
Nevertheless, as a result, 
the in-medium proton-$\sigma$ and 
neutron-$\sigma$ coupling constants, 
$g^p_\sigma (\sigma)$ and $g^n_\sigma (\sigma)$, differ from their
values in free space and the proton and neutron effective 
mass difference is reduced~\cite{saito}. This leads to a 
reduction in the  
binding energy differences below the amount one naively expects from 
the proton and neutron mass difference of about 2 MeV in free space  
(without the electromagnetic self-energy corrections) --  
see also TABLE~\ref{parameters} and FIG.~\ref{onsmass}.

\begin{figure}[hbt]
\begin{center}
\epsfig{file=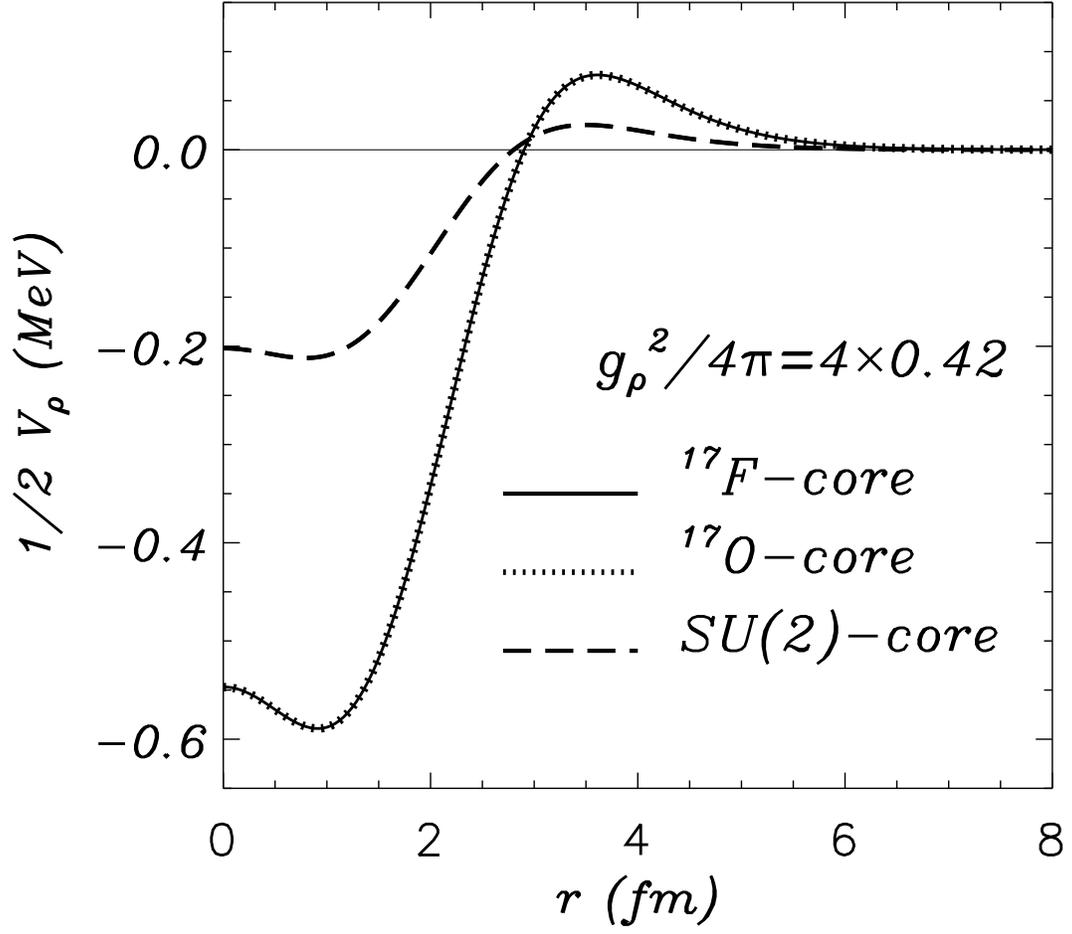,height=14cm}
\caption{Calculated $\rho$-meson (iso-vector) mean field 
potential generated by the core in  
$^{17}$O and $^{17}$F, for CSB and SU(2).}
\label{onsrho}
\end{center}
\end{figure}

The density dependence of the effective $p-n$ mass difference, which 
we have just described, is the major source of charge symmetry violation 
discussed here. On the other hand, the fact that there is a small $\rho^0$
mean field also affects the systematics as we vary $A$ and we now examine this
contribution. In FIG.~\ref{onsrho} 
we show the $\rho$-meson mean field potential 
generated by the core in $^{17}$F and $^{17}$O, for  
CSB and SU(2). There is no distinguishable difference between 
$^{17}$F and $^{17}$O for CSB.
We will evaluate the $\rho$-meson 
central and spin-orbit potential contributions 
to the single-particle energies of the valence proton and neutron 
perturbatively.
We should note that QMC gives the correct expression for the 
spin-orbit potentials, including the finite size 
of the nucleon~\cite{finite0,hyper}:
\bg
V^{s.o.}(r)\vec{l}\cdot\vec{s} &=& \frac{-1}{2m_N^{* 2}(r) r}
\left[ \Delta_\sigma
  + 3 (1 - 2\mu_s\eta (r))\Delta_\omega
+ \frac{1}{2} \tau^N_3 
(1 - 2\mu_v\eta (r))\Delta_\rho \right]\vec{l}\cdot\vec{s}.
\label{so} 
%
%
%
%
\en
We are interested in the last term in Eq.~(\ref{so}),
the $\rho$-meson spin-orbit potential, which gives opposite 
contributions for the valence proton and the valence neutron. 
Contributions from 
the other mesons to Eq.~(\ref{so}) have the same sign for protons 
and neutrons and their contributions to the binding energy differences
are therefore expected to be tiny. Furthermore, the contribution from the
effective mass difference of the proton and neutron  
is even higher order. 
Thus, we use the SU(2) value for $m^{*2}_N$ in Eq.~(\ref{so}) 
to evaluate the $\rho$-meson spin-orbit potential.
Using the calculated wave functions for the valence proton and neutron,  
obtained by solving the Dirac equations without the 
Coulomb and $\rho$-meson potentials 
or the electromagnetic self-energy corrections,
we evaluate the $\rho$-meson contributions perturbatively:
\bg
\delta E_\rho &=&  \int d^3r\, \psi_{valence}^\dagger (\vr) 
\,[\frac{1}{2} \tau^N_3 V_\rho (r)]\, \psi_{valence} (\vr), \label{erho} \\
\delta E^{s.o.}_\rho &=& \int d^3r\, \psi_{valence}^\dagger (\vr)
\,[\frac{1}{2} \tau^N_3 V_\rho^{s.o.} (r)]\, (\vec{l}\cdot\vec{s})\, 
\psi_{valence} (\vr), \label{erhoso} 
\en
where $1/2 V_\rho (r)$ is shown in FIG.~\ref{onsrho}, and 
$V_\rho^{s.o.}(r) = \frac{-1}{2m_N^{* 2}(r) r} 
(1 - 2\mu_v\eta (r))\Delta_\rho$, the third term in Eq.~(\ref{so}). 
In QMC the iso-vector magnetic moment, $\mu_v$, is calculated to be 
$\mu^{QMC}_v = 2.558$ for the bag radius $R = 0.8$ fm.  
This is somewhat smaller than the 
empirical value, $\mu^{emp.}_v = 4.7051$.
Thus, for the CSB and SU(2) calculations, we use the 
empirical value, $\mu^{emp.}_v = 4.7051$, together with the 
phenomenological coupling constant, $g^q_\rho = 4.595$, in order 
to make a more realistic estimate. 

In TABLE~\ref{summary} we summarize the calculated single-particle energies
for the valence proton and neutron of the mirror nuclei for 
two cases, CSB and SU(2).
We expect that the results for CSB are the more realistic.

Comparing the $\rho$-potential contributions for the hole states with  
core plus valence states, one notices the shell effects due to 
the $\rho$-potentials. The $\rho$-potential contributions for the 
discrepancies of the $^{15}$O -- $^{15}$N and 
$^{17}$F -- $^{17}$O binding energy differences are about  
$-$0.11 MeV and $-$0.011 MeV, respectively, while 
for the $^{39}$Ca -- $^{39}$K and 
$^{41}$Sc -- $^{41}$Ca cases, they are 
about $-$0.17 MeV and $-$0.013 MeV, respectively.
These results reflect the difference in the shell structure, 
namely hole states tend to have larger $\rho$-potential 
contributions than the core plus valence nucleon states.
This can be understood because, in the hole states, the excess 
proton or neutron sits in the region where the iso-vector density  
distribution is larger. 

For the SU(2) case, the valence state binding energy differences  
of mirror nuclei, $\delta E$, come entirely from the 
$\rho$-meson potentials. We see that $\delta E$ obtained  
in CSB is always larger than that for SU(2).
Typical values for the binding energy differences 
are a few hundred keV.

The larger binding energy differences for the valence proton and neutron
obtained in CSB indicate that the prime  
CSB effects originate in the u-d current quark
mass difference.
The resulting contribution to the binding energy differences
is of the order of about a few hundred keV. This is precisely the  
order of magnitude which is observed as
the ONS anomaly~\cite{shl,miller}. Furthermore, as we see from 
TABLE~\ref{summary}, 
the systematic dependence on $A$ is also reasonably well described,  
except for the $^{39}$Ca -- $^{39}$K case.
It is a fascinating challenge for the future to compare this result with the 
traditional mechanism involving $\rho-\omega$ mixing~\cite{blu}. This 
will involve the issue of the possible momentum dependence of the 
$\rho-\omega$ mixing amplitude \cite{GHT,miller}. In addition, one would 
need to examine whether there is any deeper connection between these 
apparently quite different sources of charge symmetry violation.

We would like to stress that the present contribution to the ONS anomaly
is based on a very simple but novel idea, namely the slight difference
between the quark scalar densities of the u and d quarks in a bound nucleon,
which stems from the u and d quark mass difference~\cite{saito}.
Our results were obtained within an
explicit shell model calculation, based on quark degrees of
freedom. They show that if charge symmetry breaking
is set through the u and d current quark mass difference so as to reproduce 
the proton and neutron masses in free space (without any
electromagnetic interaction effects), it produces 
binding energy differences for the valence
(excess) proton and neutron of mirror nuclei of a few hundred keV.
The origin of this effect within relativistic quantum mechanics is 
so simple that it is natural to conclude that a sizable fraction 
of the charge symmetry breaking in mirror nuclei arises from 
the density dependence of the u and d quark scalar densities in a bound 
nucleon.\\

\noindent
We would like to thank A. G. Williams for helpful comments on the manuscript.
This work was supported by the Australian Research Council and the Japan
Society for the Promotion of Science.


%
%
\newpage
\begin{table}[htbp]
\begin{center}
\caption{
Inputs, parameters and some of the quantities calculated in the present
study. The quantities with a star, $^*$, are those quantities calculated
at normal nuclear matter density, $\rho_0 = 0.15$ fm$^{-3}$.
The d current quark mass, $m_d$, is calculated in the 
model so as to reproduce the neutron mass, $m_n = 939.6956$ MeV, 
in free space. Phenomenological $\rho$-quark coupling constant, 
$g^q_\rho$ (phen.) ($g^2_{\rho NN}/4\pi = 4 \times 0.42$),  
the value at zero three-momentum transfer corresponding to 
Hartree approximation, is taken from TABLE 4.1 of Ref.~\protect\cite{mach}. 
}
\label{parameters}
\vspace{1em}
\begin{tabular}[hbtp]{c|cccccc}
\hline
 &$m$ (MeV) &$R$ (fm) &$B^{1/4}$ (MeV) &$z$ &$m^*$ (MeV) &$R^*$ (fm)\\
\hline
p (CSB)  &937.6423 (input)&0.8 (input)&169.81&3.305&751.928&0.7950 \\
n (CSB)  &939.6956 (input)&0.8000     &169.81&3.305&753.597&0.7951 \\
N (SU(2))&939.0 (input)   &0.8 (input)&169.97&3.295&754.542&0.7864 \\
\hline\\
 &$m_u$ (MeV) &$m_d$ (MeV) &$g^q_\sigma$ &$g^q_\omega$ &$g^q_\rho$ (QMC)
&$g^q_\rho$ (phen.)\\
\hline
CSB   & 5.0 (input)& 9.2424      &5.698 &2.744 &9.321 & 4.595\\
SU(2) & 5.0 (input)& 5.0 (input) &5.685 &2.721 &9.330 & 4.595\\
\hline
\end{tabular}
\end{center}
\end{table}
%
%
%
\begin{table}[htbp]
\begin{center}
\caption{
Calculated single-particle energies (in MeV) for $^{17}$F and $^{17}$O.
For CSB and SU(2) the phenomenological value,
$g_\rho = 4.595$ $= g^q_\rho = g_{\rho NN}$ is used, while
for Case 3, $g_\rho = 9.321$, the value determined in QMC is used.
For the valence proton and neutron the Dirac equations
are solved without including the Coulomb and $\rho$-meson
potentials or the electromagnetic self-energy corrections.
}
\label{energylevel}
\vspace{1em}
\begin{tabular}[hbtp]{r|cc|cc|cc}
\hline
 &CSB & &SU(2) & &Case 3 & \\
 &$^{17}$F &$^{17}$O &$^{17}$F &$^{17}$O &$^{17}$F &$^{17}$O \\
\hline
p 1s$_{1/2}$ &-28.800 &-28.805 &-28.663 &-28.663 &-28.991 &-28.996 \\
  1p$_{3/2}$ &-14.154 &-14.158 &-14.032 &-14.032 &-14.248 &-14.251 \\
  1p$_{1/2}$ &-12.495 &-12.499 &-12.383 &-12.383 &-12.589 &-12.592 \\
\hline
n 1s$_{1/2}$ &-33.367 &-33.372 &-32.967 &-32.967 &-33.168 &-33.173 \\
  1p$_{3/2}$ &-18.259 &-18.263 &-17.918 &-17.918 &-18.159 &-18.163 \\
  1p$_{1/2}$ &-16.587 &-16.590 &-16.258 &-16.258 &-16.487 &-16.490 \\
\hline
valence &p         &n        &p        &n        &p        &n       \\
1d$_{5/2}$ &-3.918 &-4.099  &-3.848  &-3.848  &-3.918  &-4.100 \\
\hline
\end{tabular}
\end{center}
\end{table}
%
%
%
\newpage
\begin{table}[htbp]
\begin{center}
\caption{
Calculated single-particle energies (in MeV) of mirror nuclei.
$\delta E_\rho$, and $\delta E_\rho^{s.o.}$ stand for the contributions from  
the $\rho$-meson central and spin-orbit potentials of the core nucleus, 
respectively. (See also Eqs.~(\protect\ref{erho}) and
(\protect\ref{erhoso}).)
The valence proton or neutron is indicated inside brackets.
The discrepancies between the experimental 
values and the theoretical expectations in the absence of 
charge symmetry violating strong interactions, 
are taken from TABLE II of Ref~\protect\cite{shah}, by averaging 
over the theoretical values. 
For the other explanations see the caption of 
TABLE~\protect\ref{energylevel}.
}
\label{summary}
\vspace{1em}
\begin{tabular}[hbtp]{c|cc|cc}
\hline
 &CSB  & &SU(2) & \\
\hline
 & & & & \\
 &$^{15}$O(p)&$^{15}$N(n)&$^{15}$O(p)&$^{15}$N(n)\\
\hline
1p$_{3/2}$      &-14.397 &-14.631 &-14.306 &-14.306 \\
$\delta E_\rho$        &-0.046  &0.047   &-0.040  &0.040   \\
$\delta E_\rho^{s.o.}$ &-0.009  &0.009   &-0.008  &0.008   \\
Total           &-14.452 &-14.575 &-14.353 &-14.258 \\
\hline
$\delta E=E(p)-E(n)$ 
 &$\delta E =$&0.123 &$\delta E$ =&-0.095 \\
 &observed =&0.227   &            &       \\
\hline
 & & & & \\
 &$^{17}$F(p)&$^{17}$O(n)&$^{17}$F(p)&$^{17}$O(n)\\
\hline
1d$_{5/2}$      &-3.918  &-4.099  &-3.848  &-3.848  \\
$\delta E_\rho$        &-0.011  &0.011   &-0.009  &0.009   \\
$\delta E_\rho^{s.o.}$ &0.006   &-0.006  &0.005   &-0.005  \\
Total           &-3.923  &-4.094  &-3.852  &-3.843  \\
\hline
\hline
$\delta E=E(p)-E(n)$ 
 &$\delta E =$&0.171 &$\delta E$ =&-0.009 \\
 &observed =&0.218   &            &       \\
\hline
 & & & & \\
 &$^{39}$Ca(p)&$^{39}$K(n)&$^{39}$Ca(p)&$^{39}$K(n)\\
\hline
1d$_{3/2}$      &-16.407 &-16.689 &-16.332 &-16.332 \\
$\delta E_\rho$        &-0.071  &0.072   &-0.065  &0.065   \\
$\delta E_\rho^{s.o.}$ &-0.016  &0.016   &-0.015  &0.015   \\
Total           &-16.493 &-16.601 &-16.411 &-16.252 \\
\hline
$\delta E=E(p)-E(n)$ 
 &$\delta E =$&0.108 &$\delta E$ =&-0.159 \\
 &observed =&0.340   &            &       \\
\hline
 & & & & \\
&$^{41}$Sc(p)&$^{41}$Ca(n)&$^{41}$Sc(p)&$^{41}$Ca(n)\\
\hline
1f$_{7/2}$      &-6.970  &-7.210  &-6.900  &-6.900  \\
$\delta E_\rho$        &-0.018  &0.018   &-0.016  &0.016   \\
$\delta E_\rho^{s.o.}$ &0.012   &-0.012  &0.011   &-0.011  \\
Total           &-6.976  &-7.204  &-6.905  &-6.894  \\
\hline
$\delta E=E(p)-E(n)$ 
 &$\delta E =$&0.228 &$\delta E$ =&-0.011  \\
 &observed =&0.463   &            &       \\
\hline
%
\end{tabular}
\end{center}
\end{table}
%
%
%
\newpage
%
%
%
\newpage
%
%
\end{document}